\documentclass[12pt, one column]{IEEEtran}
\usepackage{amsmath, amsthm, amsfonts}
\usepackage[english]{babel}
\usepackage{graphicx}
\usepackage{tabularx}
\usepackage{multirow}
\usepackage{enumerate}
\usepackage{setspace}
\usepackage[ruled,vlined]{algorithm2e}
\begin{document}

\title{Hardware Implementation of Compressed Sensing based Low Complex Video Encoder}

\author{{B.K.N.Srinivasarao, Indrajit Chakrabarti\\
Department of Electronics and Electrical Communication Engineering\\
Indian Institute of Technology, Kharagpur, INDIA\\
E.Mail : srinu.bkn@iitkgp.ac.in,
indrajit@ece.iitkgp.ernet.in}}
\maketitle
\thispagestyle{empty}
\doublespacing
\begin{abstract}

This paper presents a memory efficient VLSI architecture of low complex video encoder using three dimensional (3-D) wavelet and Compressed Sensing (CS) is proposed for space and low power video applications. Majority of the conventional video coding schemes are based on hybrid model, which requires complex operations like transform coding (DCT), motion estimation and deblocking filter at the encoder. Complexity of the proposed encoder is reduced by replacing those complex operations by 3-D DWT and CS at the encoder. The proposed architecture uses 3-D DWT to enable the scalability with levels of wavelet decomposition and also to exploit the spatial and the temporal redundancies. CS provides the good error resilience and coding efficiency. At the first stage of the proposed architecture for encoder, 3-D DWT has been applied (Lifting based 2-D DWT in spatial domain and Haar wavelet in temporal domain) on each frame of the group of frames (GOF), and in the second stage CS module exploits the sparsity of the wavelet coefficients. Small set of linear measurements are extracted by projecting the sparse 3-D wavelet coefficients onto random Bernoulli matrix at the encoder. Compared with the best existing 3-D DWT architectures, the proposed architecture for 3-D DWT requires less memory and provide high throughput. For an N$\times$N image, the proposed 3-D DWT architecture consumes a total of only $2*(3N + 40P)$ words of on-chip memory for the one level of decomposition. The proposed architecture for an encoder is first of its kind and to the best of my knowledge, no architecture is noted for comparison. The proposed VLSI architecture of the encoder has been  synthesized on 90-nm CMOS process technology and results show that it consumes 90.08 mW power and  occupies an area equivalent to 416.799 K equivalent gate at frequency of 158 MHz.  The proposed architecture has also been synthesised for the Xilinx zync 7020 series field programmable gate array (FPGA).

\end{abstract}


\section{Introduction}
\par Current video coding standards (e.g.,H.264 and HEVC)\cite{H.264}\cite{HEVC} are able to provide good compression using a high-complexity encoders. At the encoder, motion estimation (using block-matching) has been applied between adjacent frames to exploit the temporal redundancy. Then each reference and residual frame (motion-compensated differences)  is divided in to non overlapping blocks (block size may vary from 8$\times$8 to 64$\times$64 pixels) and apply the transform coding on each block (e.g., DCT) to exploit the spatial redundancy. Motion estimation and transform coding accounts for nearly 70$\%$ of the total complexity of the encoder \cite{Richardson}. Moreover, block wise transform coding leads to blocking artifacts in the motion compensated frame and it may reduced by using deblocking filter. However, which may further increase the complexity of the encoder. In contrast, the decoder complexity is very low. The main function of the decoder is to reconstruct the video frames by using reference frame, motion-compensated residuals and motion vectors. They are more suitable for the broadcasting applications, where a high complexity encoder would support thousands of low complex decoders. However, conventional video coding schemes are not suitable for applications requires low complexity encoders like mobile phones and camcorders. There requires low complex, low power and low cost devices. High complex encoder enables increase in compression ratio and power consumption. Therefore, to increase battery life in mobile devices, a low-complexity encoder with good coding efficiency is highly desirable.

\par In a mobile video broadcast network (wireless networks), a video source is broadcast to multiple receivers and may have various channel capacities, display resolutions, or computing facilities. It is necessary to encode and transmit the video source once, but allow any subset of the bit stream to be successfully decoded by a receiver. In order to reduce the error rate in wireless broadcast network, error correction coding such as Reed-Solomon (RS) code and convolutional code has been widely used. However, this type of channel coding is not flexible. It can correct the bit errors only if the error rate is smaller than a given threshold. Therefore, it is hard to find a single channel code suitable for different channels having different capacities. For broadcast applications, without the feedback from individual receivers, the sender can re-transmit data that are helpful to all the receivers. These requirements are indeed difficult and challenging for traditional channel coding design.
From the above requirements it is desired to have a encoder with less complex, good coding efficiency, error resilience, scalable and support the realtime application. 

\par
\begin{figure} 
\centering
\includegraphics [height=100mm,width=80mm]{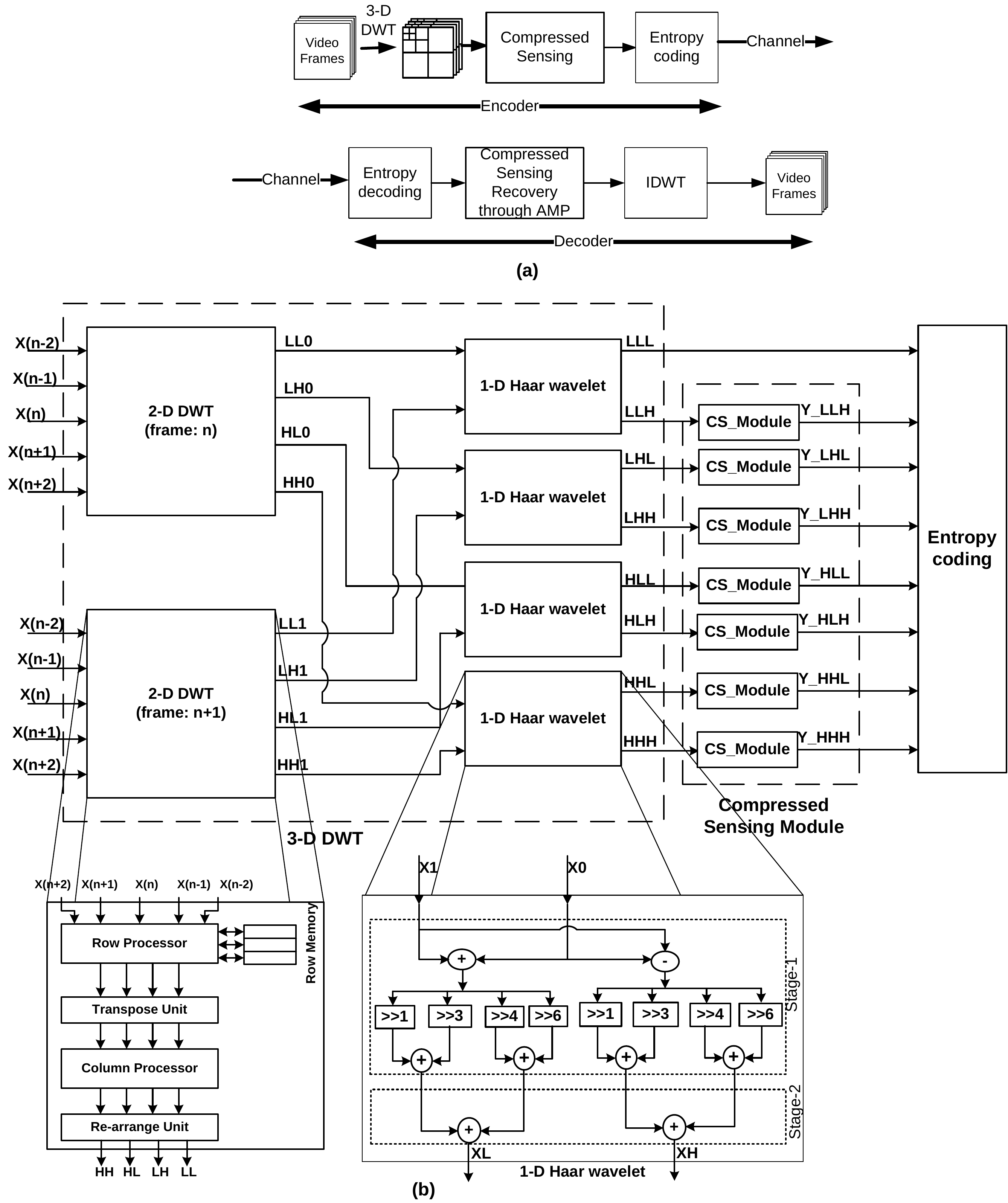}
\caption {(a) Block diagram for CS based Scalable Video Codec (b)Detailed block diagram of Encoder}
\label{blockdia_1}
\end{figure}

 This paper introduces a new VLSI architecture for scalable low complex encoder using 3-D DWT and compressed sensing. Fig.~\ref{blockdia_1}(a) shows the block diagram of low complex video codec (encoder and decoder). Encoder has 3-D DWT and CS as main functional modules shown in Fig.~\ref{blockdia_1}(b). 3-D DWT module provides the scalability with the levels of decomposition and also exploit the spatial and temporal redundancies of the video frames. 3-D DWT module of the encoder replaces the transform coding, motion estimation and deblocking filters of the current video coding system. CS module utilize the sparse nature of the wavelet coefficients and projects on the random Bernoulli matrices for selecting the measurements at the encoder to enable the compression and approximate message passing algorithm for reconstruction at the decoder. CS module provides the good compression ratio and improves the error resilience.  As a result the proposed architecture enjoys lesser complexity at the encoder and marginal complexity at the decoder. \\
 
\par From the last two decades, several hardware designs have been noted for implementation of 2-D DWT and 3-D DWT for different applications. Majority of the designs are developed based on three categories, viz. (i) convolution based (ii) lifting-based and (iii) B-Spline based.  Most of the existing architectures are facing the difficulty with larger memory requirement, lower throughput, and complex control circuit. In general the circuit complexity is denoted by two major components viz, arithmetic and Memory component. Arithmetic component includes adders and multipliers, whereas memory component consists of temporal memory and transpose memory. Complexity of the arithmetic components is fully depends on the DWT filter length. In contrast size of the memory component  is depends on dimensions of the image. As image resolutions are continuously increasing (HD to UHD), image dimensions are very high compared to filter length of the DWT, as a result complexity of the memory component occupied major share in the overall complexity of DWT architecture.\\

\par Convolution based implementations \cite{3D_conv_dai}-\cite{3D_conv_mohanty} provides the outputs within less time but require high amount of arithmetic resources, memory intensive and occupy larger area to implement. Lifting based a implementations requires less memory, less arithmetic complex and possibility to implement in parallel. However it require long critical path, recently huge number of contributions are noted to reduce the critical path in lifting based implementations. For a general lifting based structure \cite{dwt_1} provides critical path of 4$T_{m}+8T_{a}$, by introducing 4 stage pipeline it cut down to $ T_{m}+2T_{a}$. In \cite{3D_flip} Huang et al., introduced a flipping structure it further reduced the critical path to $ T_{m}+T_{a}$. Though, it reduced the critical path delay in lifting based implementation, it requires to improve the memory efficiency. Majority of the designs which implement the 2-D DWT,  first by applying 1-D DWT in row-wise and then apply 1-D DWT in column wise. It require huge amount of memory to store these intermediate coefficients. To reduce this memory requirements, several DWT architecture have been proposed by using line based scanning methods \cite{3D_huang}-\cite{3D_xiong}. Huang et al., \cite{3D_huang}-\cite{3D_huang2} given brief details of B-Spline based 2-D IDWT implementation and discussed the memory requirements for different scan techniques and also proposed a efficient overlapped strip-based scanning to reduce the internal memory size. Several parallel architectures were proposed for lifting-based 2-D DWT \cite{3D_huang2}-\cite{3D_yusong2}. Y. Hu et al. \cite{3D_yusong2}, proposed a modified strip based scanning and parallel architecture for 2-D DWT is the best memory-efficient design among the existing 2-D DWT architectures, it requires only 3N + 24P of on chip memory for a N$\times$N image with $ P $ parallel processing units (PU). Several lifting based 3-D DWT architectures are noted in the literature \cite{3D_zheng}-\cite{3D_darji} to reduce the critical path of the 1-D DWT architecture and to decrease the memory requirement of the 3-D architecture. Among the best existing designs of 3-D DWT, Darji et al. \cite{3D_darji} produced best results by reducing the memory requirements and gives the throughput of 4 results/cycle. Still it requires the $4N^{2}+10N$ on-chip memory. 
 
\par Based on the ideas of compressed sensing (CS) \cite{CS_11}-\cite{CS_13}, several new video codecs \cite{CS_10}-\cite{CS_plonka} have been proposed in the last few years. Wakin \textit{et al.} \cite{CS_10} have introduced the compressive imaging and video encoding through single pixel camera. From his research results, Wakin has established that 3-D wavelet transform is a better choice for video compared to 2-D (two-dimensional) wavelet transform. Y. Hou and F. Liu \cite{32} have proposed a system of low complexity, where sparsity extracted is from residuals of successive non-key frames and CS is applied on those frames. Key frames are fully sampled resulting in increased bit-rate. Moreover, performing motion estimation and compensation while predicting the non key frames increases the encoder complexity. S. Xiang and Lin Cai \cite{33} proposed a CS based scalable video coding, in which the base layer is composed of a small set of DCT coefficients while the enhancement layer is composed of compressed sensed measurements. It uses DCT for I frames and undecimated DWT (UDWT) for CS measurements which increases the complexity at the decoder to a great extent.  Jiang \textit{et al.} \cite{31} proposed CS based scalable video coding using total variation of the coefficients of temporal DCT. Scalability is enabled by  multi-resolution measurements while the video signal is reconstructed by total variation minimization by augmented Lagrangian and alternating direction algorithms (TVAL3) \cite{tval3} at the decoder. However, it increases the decoder complexity, making hardware implementation quite difficult.  J. Ma \textit{et al.} \cite{CS_plonka} introduced the fast and simple on-line based encoding and decoding by forward and backward splitting algorithm. Though encoder complexity is low, scalability is not achieved and decoder complexity is very high. Most of the recently proposed video codecs \cite{CS_10}-\cite{CS_plonka}, which are assumed to be of uniform sparsity, are available for all the video frames and a fixed number of measurements are transmitted to decoder for all the frames. Depending on the content of the video frame, sparsity may change. A fixed number of measurements may cause an increase in bit-rate (decrease in compression ratio). \\

\par This paper introduces a new compressed sensing based low complex encoder architecture using 3-D DWT. The proposed method uses the random Bernoulli sequence at the encoder for selecting the measurements and the approximate message passing algorithm for reconstruction at the decoder. Major contributions of the present work may be stated as follows. Firstly the proposed framework has revised the MCTF based SVC \cite{CS_4} model by introducing compressed sensing concepts to increase the compression ratio and to reduce the complexity. As a result, the proposed framework ensures low complexity at the encoder and marginal complexity at the decoder. Secondly, we proposed a new architecture for 3-D DWT, which requires only  $2*(3N + 40P)$ words of on-chip memory with a throughput of 8 results/cycle. Thirdly, we proposed a efficient architecture for compressed sensing module. 

\par Organization of the paper as follows. Fundamentals of 3-D DWT and compressed sensing is presented in Section II. Detailed description of the proposed architecture for 3-D DWT  and compressed sensing modules are provided in section III and IV respectively . Results and comparison are given in Section V. Finally, concluding remarks are given in Section VI.

\section{Theoretical Framework}

This section presents a theoretical background of the wavelets and compressed sensing. 3-D DWT has been used to exploit the spatial and temporal redundancies of the video, thereby it eliminates the complex operations like ME, MC and deblocking filter. Compressed sensing is used to provide the error resilience and coding efficiency.

\subsection{Discrete Wavelet Transform}
\begin{figure} 
\centering
\includegraphics [height=50mm,width=40mm]{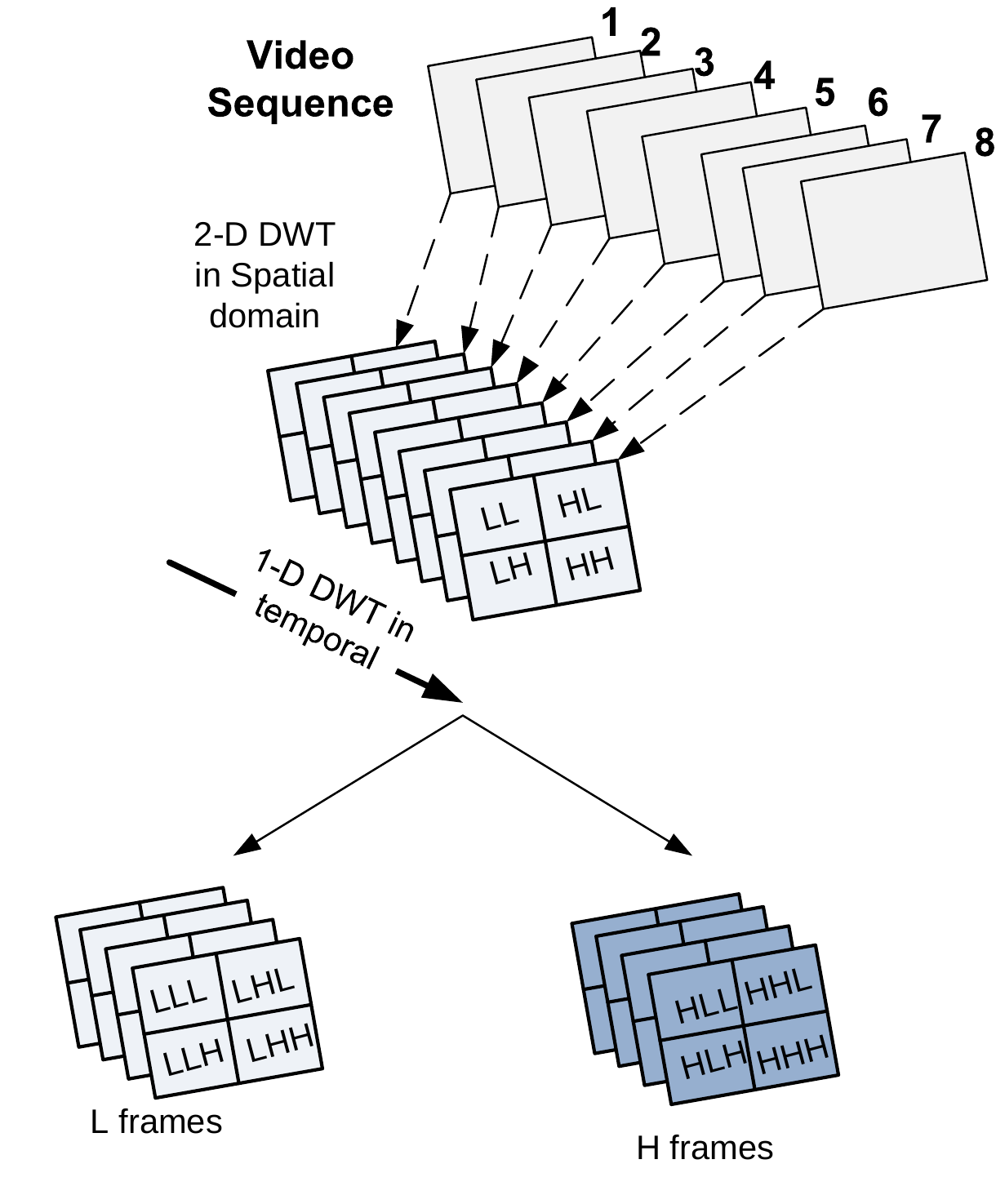}
\caption {3-D wavelet by combining 2-D spatial and 1-D temporal}
\label{3dDWT}
\end{figure}
\par Lifting based wavelet transform designed by using a series of matrix decomposition specified by the Daubechies and Sweledens in \cite{dwt_1}. By applying the  flipping \cite{3D_flip} to the lifting scheme, the multipliers in the longest delay path are eliminated, resulting in a shorter critical path. The original data on which DWT is applied is denoted by $X[n]$; and the 1-D DWT outputs are the detail coefficients $H[n]$ and approximation coefficients $L[n]$.  For the Image (2-D) above process is performed in rows and columns as well.  Eqns.(1)-(6) are the design equations for flipping based lifting (9/7) 1-D DWT \cite{3D_flip2} and the same equations are used to implement the proposed row processor (1-D DWT) and column processor (1-D DWT). 
\begin{figure}[H]
\label{lift_2d_eq}
\begin{align}
{H_1}[n] &\leftarrow a'*X[2n-1]+\{X[2n]+X[2n-2]\}  \ldots P1\\
{L_1}[n]&\leftarrow b'*X[2n]+\{{H_1}[n]+{H_1}[n-1]\}  \ldots U1\\
{H_2}[n] &\leftarrow c'*{H_1}[n]+\{{L_1}[n]+{L_1}[n-1]\}\ldots P2\\
{L_2}[n] &\leftarrow d'*{L_1}[n]+\{{H_2}[n]+{H_2}[n-1]\}\ldots U2\\
H[n] &\leftarrow K0* \{{H_2}[n]\}\\
L[n] &\leftarrow K1* \{{L_2}[n]\}
\end{align}
\end{figure}
Where $ a'=1/\alpha $, $ b'=1/\alpha\beta $, $ c'=1/\beta\gamma $, $ d'=1/\gamma\delta $, $ K0= \alpha\beta\gamma/\zeta$, and $ K1= \alpha\beta\gamma\delta\zeta$  \cite{dwt_1}. The lifting step coefficients $ \alpha$, $ \beta$, $\gamma $, $ \delta $ and scaling coefficient $\zeta $ are constants and its values  $ \alpha = -1.586134342$, $ \beta =-0.052980118$, $\gamma =0.8829110762$, and $ \delta =0.4435068522$, and  $\zeta  = 1.149604398.$

Lifting based wavelets are always memory efficient and easy to implement in hardware. The lifting scheme consists of three steps to decompose the samples, namely, splitting, predicting (eqn. (1) and (3)), and updating (eqn. (2) and (4)). 

Haar wavelet transform is orthogonal and simple to construct and provide fast output. By considering the advantages of the Haar wavelets, the proposed architecture uses the Haar wavelet to perform the 1-D DWT in temporal direction (between two adjacent frames). Sweldens \textit{et al.} \cite{daub_Haar} developed a lifting based Haar wavelet. The equations of the lifting scheme for the Haar wavelet transform is as shown in eqn.(\ref{eq2})
\begin{equation}
\label{eq2}
\left[ \begin{array}{l}
L\\
H
\end{array} \right] = \left( {\begin{array}{*{20}{c}}
{\sqrt 2 }&0\\
0&{\frac{1}{{\sqrt 2 }}}
\end{array}} \right)\left( {\begin{array}{*{20}{c}}
1&{S(z)}\\
0&1
\end{array}} \right)\left( {\begin{array}{*{20}{c}}
1&0\\
{ - P(z)}&1
\end{array}} \right)\left( \begin{array}{l}
{X_0}(z)\\
{X_1}(z)
\end{array} \right)
\end{equation}
\begin{equation}
\label{eq3}
\begin{array}{l}
L = {\textstyle{1 \over {\sqrt 2 }}}({X_0} + {X_1})\\
H = {\textstyle{1 \over {\sqrt 2 }}}({X_1} - {X_0})
\end{array}
\end{equation}
Eqn.(\ref{eq3}) is extracted by substituting Predict value $P(z)$ as 1 and Update step $ S(z) $ value as 1/2 in eqn.(\ref{eq2}), which is used to develop the temporal processor to apply 1-D DWT in temporal direction ($ 3^{rd} $ dimension). Where L and H are the low and High frequency coefficients respectively.
\par The process which is shown in Fig. \ref{3dDWT} represents the one level decomposition in spatial and temporal. Among all the sub-bands, only LLL sub-band (LL band of L-frames) is fully sampled and transmitted without applying any CS techniques because it represents the image in low resolution (Base layer in SVC domain) which is not sparse. All the other sub-bands (3-D wavelet coefficients) except LLL exhibit approximate sparsity (Near to zero) and hard thresholding has been applied (consider as zero if value is less than threshold). After this step, conventional encoders use EZW coding to encode these wavelet coefficients which is complex to implement in hardware. EZW coding is replaced by CS in the proposed framework which exploits the sparsity preserving nature of random Bernoulli matrix by projecting the wavelet coefficients onto them. DWT version of each frame consists of four sub-bands. All the LL sub-bands of L-frames have large wavelet coefficients. Remaining three bands of L-frames and four sub-bands of H-frames exhibits sparsity on which compressed sensing is applied. 

\subsection{Compressed Sensing}
Compressed sensing is an innovative scheme that enables sampling below the Nyquist rate, without (or with small) drop in reconstruction quality. The basic principle behind the compressed sensing consists in exploiting sparsity of the signal in some domain. In the proposed work, CS has been applied in wavelet domain.\\
Let $x = \lbrace x[1], . . . ,x[N]\rbrace$ be a set of $ N $ real and discrete-time samples. Let s be the representation of $ x $ in the $ \Psi $ (transform) domain, that is:
\begin{equation}
x = \Psi s = \sum\limits_{i = 1}^N {\Psi _i}{{s_i}} 
\end{equation}
where $ s $ = $  [s_1, . . . , s_N]$ is a weighted coefficients vector, $ s_i $ = $\left\langle {x,{\Psi _i}} \right\rangle$, and $\Psi  = [{\Psi _{1,}}|{\Psi _{2,}}|....|{\Psi _N}]$ is an $N \times N$ basic matrix. Assume that the vector $ x $ is $ K $-sparse ($ K $ coefficients of $ s $ are non-zero) in the domain $ \Psi $, and $ K \ll N $. To get the sparsity of the signal $ x $, conventional transform coding is applied on whole signal $ x $ (all $ N $ samples) by using $ s = \Psi^{T}x $ and gives the $ N $ transform coefficients. Among the N coefficients, $N-K$ or more coefficients are discarded because they carry negligible energy and the remaining are encoded. The basic idea of CS is to remove this ``sampling redundancy'' by taking only $ M $ samples of the signal, where $ K < M  \ll N. $ Let $ y $ be an $ M $-element measurement
vector given by: $y = \Phi x$ or $  y = \Phi \Psi s$ with $y \in \mathbb{R}^{M}$, $\Phi \in \mathbb{R}^{M\times N} $ are non-adaptive linear projections of a signal $x$ $\in \mathbb{R}^{N}$ with typically  $M \ll N$. 

Recovering the original signal $ x $ means solving an under-determined linear equation with usually no unique solution. However, the signal $ x $ can be recovered losslessly from $M \ge K$ measurements, if the measurement matrix $ \Phi $ is designed in such a way that, it should preserve the geometry of the sparse signals and each of its $M \times K$ sub-matrices possesses full rank. This property is called Restricted Isometry Property (RIP) and
mathematically, it ensures that $\|x1-x2\|_2 \approx\|\Phi x1-\Phi
x2\|_2$. Where $\|y\|_2$ represents the  $ \ell_2 $- norm of the vector $y $. It has been observed that the random matrices drawn from independent and identically distributed (i.i.d.) Gaussian or Bernoulli distributions satisfy the RIP property with high probability. 

\par The problem of signal recovery from CS measurements is very well studied in the recent years and there exists a host of algorithms that have been proposed such as Orthogonal Matching
Pursuit (OMP) \cite{26}-\cite{28}, Iterative Hard-Thresholding (IHT) \cite{IHT}, Iterative
Soft-Thresholding (IST) \cite{IST}. Although recently introduced Approximate Message Passing (AMP) algorithm \cite{dono_1} shows a similar structure to IHT and IST, it exhibits faster convergence. Literature \cite{dono_1},\cite{dono_2} shows that AMP performs excellently for many deterministic and highly structured matrices.\\

\section{Proposed architecture for 3-D DWT}

The proposed architecture for 3-D DWT comprising of two parallel spatial processors (2-D DWT) and four  temporal processors (1-D DWT), is depicted in Fig. \ref{blockdia_1}(b).  After applying 2-D DWT on two consecutive frames, each spatial processor (SP) produces 4 sub-bands, viz. LL, HL, LH and HH and are fed to the inputs of four temporal processors (TPs) to perform the temporal transform. Output of these TPs is a low frequency frame (L-frame) and a high frequency frame (H-frame). Architectural details of the spatial processor and temporal processors are discussed in the following sections.

\subsection{Architecture for Spatial Processor}
\begin{figure} 
\centering
\includegraphics [height=100mm,width=100mm]{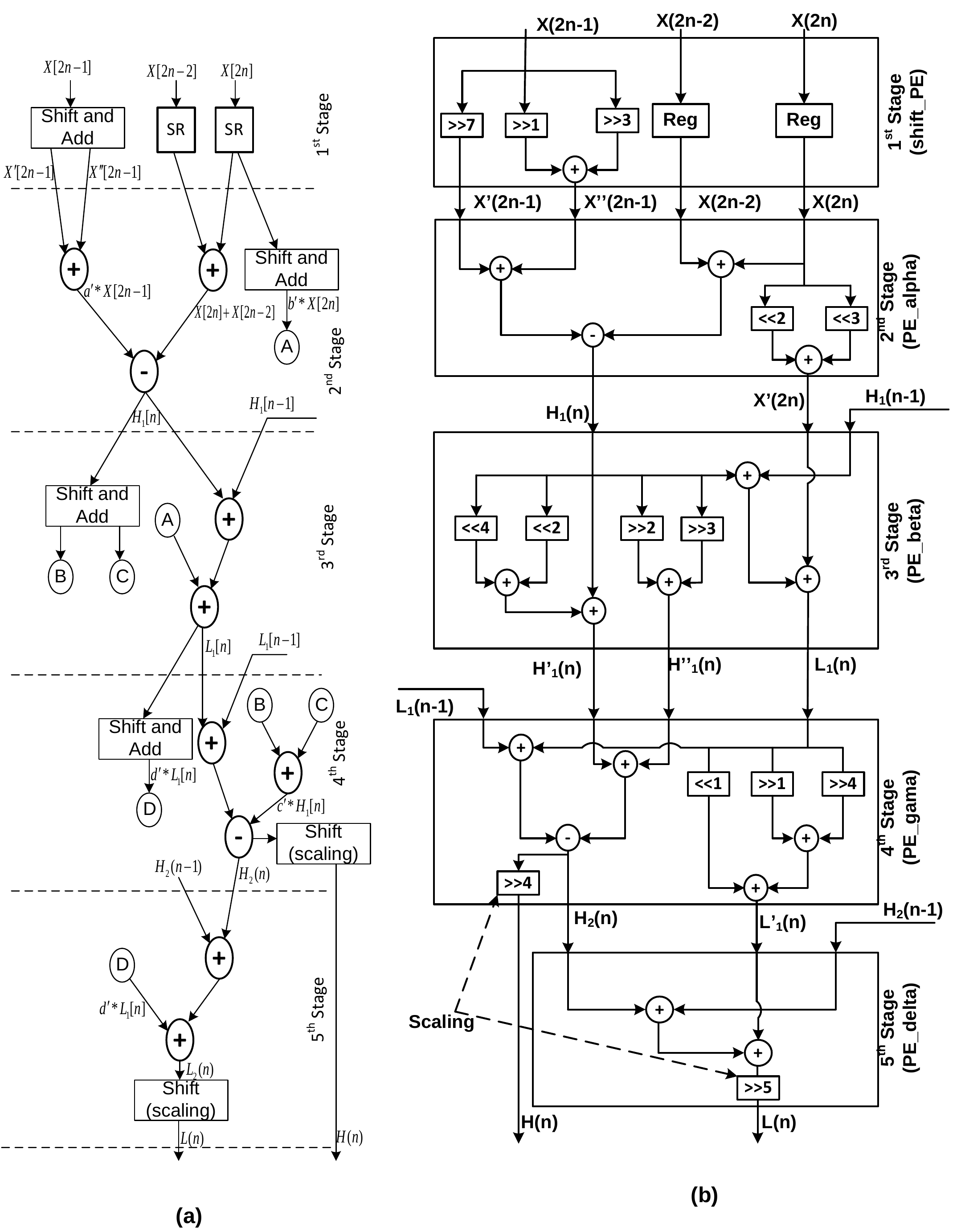}
\caption{(a) Data Flow Graph of modified 1-D DWT architecture (b)Structure of Processing Unit }
\label{3d_2}
\end{figure}

In this section, we propose a new parallel and memory efficient lifting based 2-D DWT architecture denoted by spatial processor (SP) and it consists of row and column processors. The proposed SP is a revised version of the architecture developed by the Y. Hu et al.\cite{3D_yusong2}. The proposed architecture utilizes the strip based scanning \cite{3D_yusong2} to enable the trade-off between external memory and internal memory. To reduce the critical path in each stage flipping model \cite{3D_flip}-\cite{3D_flip2} is used to develop the processing element (PE). Each PE has been developed with shift and add techniques in place of multiplier. Lifting based (9/7) 1-D DWT process has been performed by the processing unit (PU) in the proposed architecture. To reduce the CPD, processing unit is designed with five pipeline stages and multipliers are replaced with shift and add techniques. This modified PU reduces the CPD to $ 2T_{a} $ (two adder delay). Fig.~\ref{3d_2}(a) shows the data flow graph (DFG) of the proposed PU and Fig.~\ref{3d_2}(b) depicts the internal architecture of the proposed PU. The number of inputs to the spatial processor is equal to 2P+1, which is also equal to the width of the strip. Where P is the number of parallel processing units (PUs) in the row processor as well as column processor. We have designed the proposed architecture with two parallel processing units (P = 2). The same structure can be extended to P = 4, 8, 16 or 32 depending on external bandwidth. Whenever row processor produces the intermediate results, immediately column processor start to process on those intermediate results. Row processor takes 5 clocks to produce the temporary results then after column processor takes 5 more clocks to to give the 2-D DWT output; finally, temporal processor takes 2 more clock after 2-D DWT results are available to produce 3-D DWT output. As a summary, proposed 2-D DWT and 3-D DWT  architectures have constant latency of 10 and 12 clock cycles respectively, regardless of image size N and number of parallel PUs (P). Details of the row processor and column processor are given in the following sub-sections.

\begin{figure*} 
\centering
\includegraphics [height=120mm,width=130mm]{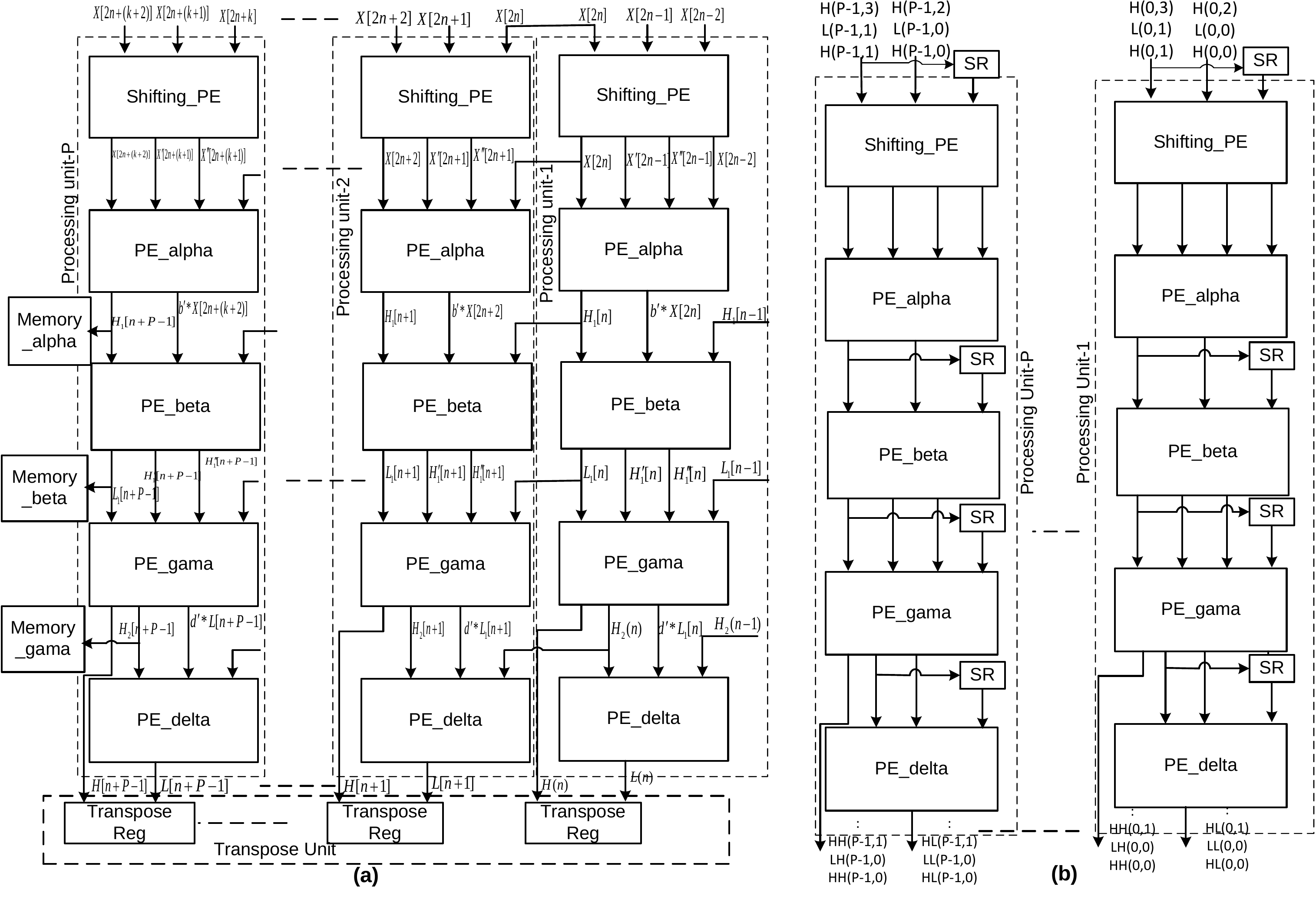}
\caption{(a)Row Processor  (b) Column Processor}
\label{3d_1}
\end{figure*}

\begin{figure} 
\centering
\includegraphics [height=70mm,width=50mm]{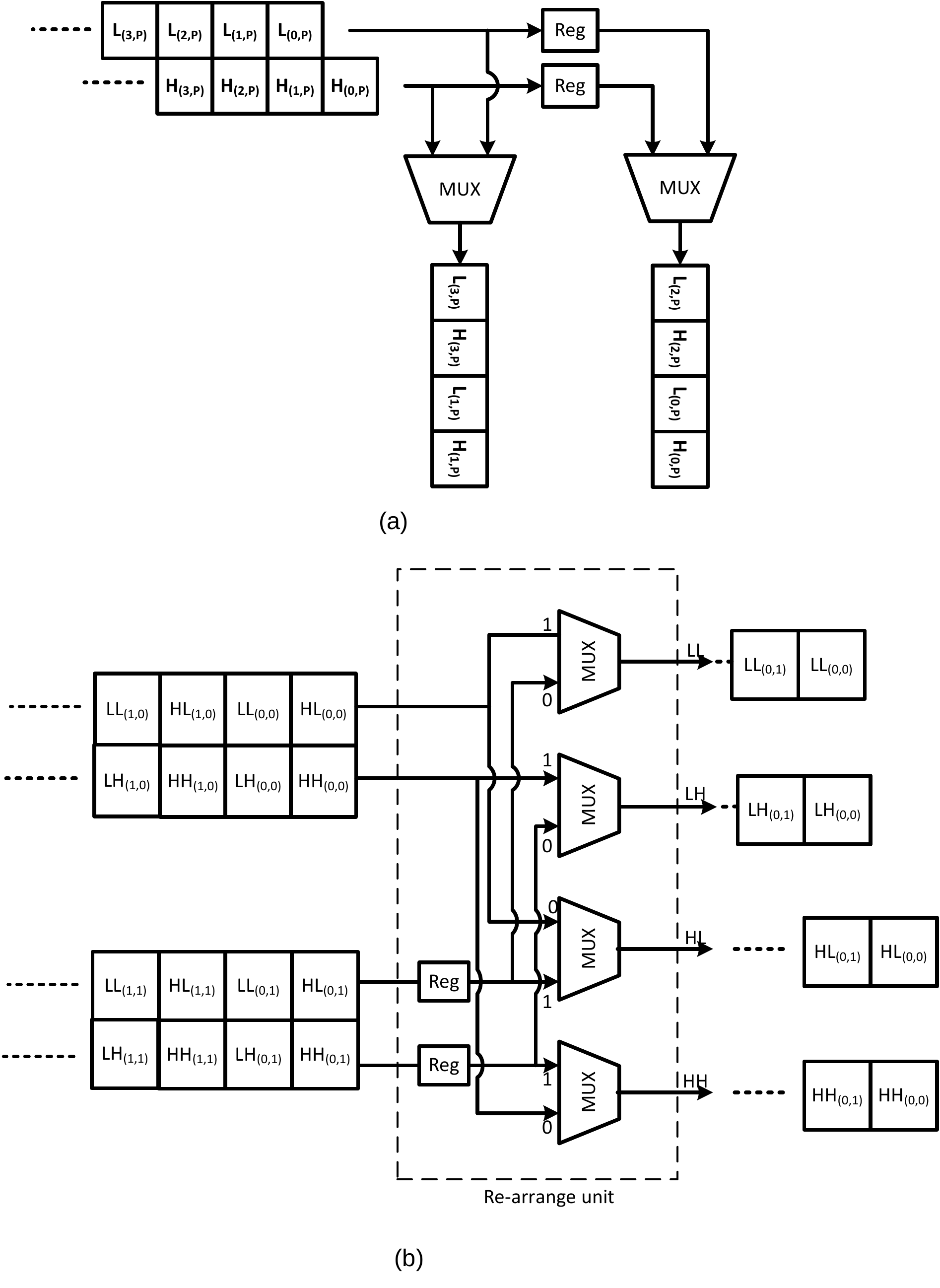}
\caption{(a) Transpose Register (Ref:\cite{3D_yusong2}) (b) Re-arrange Unit}
\label{3d_3}
\end{figure}

\subsubsection{Row Processor (RP)}
Let X be the image of size N$ \times $N, extend this image by one column by using symmetric extension. Now image size is N$\times $(N+1). Refer \cite{3D_yusong2} for the structure of strip based scanning method. The proposed architecture initiates the DWT process in row wise through row processor (RP) then process the column DWT by column processor (CP).  Fig.~\ref{3d_1}(a).  shows the generalized structure for a row processor with P number of PUs. P = 2 has been considered for our proposed design.  For the first clock cycle, RP get the pixels from X(0,0) to X(0,2P) simultaneously. For the second clock RP gets the pixels from next row i.e. X(1,0) to X(1,2P), the same procedure continues for each clock till it reaches the bottom row i.e., X(N,0) to X(N,2P). Then it goes to the next strip and RP get the pixels from X(0,2P) to X(0,4P) and it continues this procedure for entire image. Each PU consists of five pipeline stages and each pipeline stage is processed by one processing element (PE) as depicted in Fig.~\ref{3d_2}(b). First stage (shift$\_$PE) provide the partial results which is required at $2^{nd}$ stage (PE$\_$alpha), likewise Processing elements PE$\_$alpha to PE$\_$delta ($2^{nd}$stage  to $5^{th}$ stage) gives the partial results along with their original outputs. i.e. consider the PU-1, PE$\_$alpha needs to provide output corresponding to eqn.(1) ($ H_{1}[n]$), along with $H_{1}[n] $, it also provides the partial output $ X'[2n]$ which is required for the PE$\_$beta. Structure of the PEs are given in the Fig.~\ref{3d_2}(b), it shows that multiplication is replaced with the shift and add technique. The original multiplication factor and the value through the shift and add circuit are noted in Table.\ref{tab1}, it shows that variation between original and adopted one is extremely small. The maximum CPD provided by the these PEs is $ 2T_{a} $. The outputs $ H_{1}[n+P-1]$, $ L_{1}[n+P-1]$, and $ H_{2}[n+P-1]$ corresponding to PE$\_$alpha and PE$\_$beta of last PU and PE$\_$gama of last PU is saved in the memories Memory$\_$alpha, Memory$\_$beta  and Memory$\_$gama respectively. Those stored outputs are inputted for next subsequent columns of the same row. For a N$\times $N image rows is equivalent to N. So the size of the each memory is N$\times $1 words and total row  memory to store these outputs is equals to 3N. Output of each PU are under gone through a process of scaling before it producing the outputs H and L. These outputs are fed to the transposing unit. The transpose unit has P number of transpose registers (one for each PU). Fig.~\ref{3d_3}(a) shows the structure of transpose register, and it gives the two H and two L data alternatively to the column processor. 

\subsubsection{Column Processor (CP)}

The structure of the Column Processor (CP) is shown in Fig.~\ref{3d_1}(b). To match with the RP throughput, CP is also designed with  two number of PUs in our architecture. Each transpose register produces a pair of H and L in an alternative order and are fed to the inputs of one PU of the CP. The partial results produced are consumed by the next PE after two clock cycles. As such, shift registers of length two are needed within the CP between each pipeline stages for caching the partial results (except between $ 1^{st} $ and $2^{nd} $ pipeline stages). At the output of the CP, four sub-bands are generated in an interleaved pattern, i.e., (HL,HH), (LL,LH), (HL,HH), (LL,LH), and so on. Outputs of the CP are fed to the re-arrange unit. Fig.~\ref{3d_3}(b) shows the architecture for re-arrange unit, and it provides the outputs in sub-band order i.e LL, LH, HL and HH simultaneously, by using P registers and 2P multiplexers. For multilevel decomposition, the same DWT core can be used in a folded architecture with an external frame buffer for the LL sub-band coefficients.
\begin{table}[]
\centering
\caption{Original and adopted values for multiplication}
\label{tab1}
\centering
\begin{tabular}{|l|l|l|}
\hline
     & Original  & Multiplier \\
PE       &  Multiplier &  value through  \\
     &  Value & shift and add \\
\hline
PE$ \_ $alpha & $ a'$=-0.6305      & $ a'$=-0.6328 \\
\hline
PE$ \_ $beta  & $ b'$=11.90        &$ b'$=12      \\
\hline
PE$ \_ $gama  &$ c'$=-21.378      & $ c'$=-21.375 \\
\hline
PE$ \_ $delta & $ d'$=2.55         & $ d'$=2.565  \\
\hline
\end{tabular}
\end{table}   

\subsection{Architecture for Temporal Processor (TP)}
Eqn.(\ref{eq3}) shows that Haar wavelet transform depends on two adjacent pixels values. As soon as spatial processors are provide the 2-D DWT results, temporal processors starts processing on the spatial processor outputs (2-D DWT results) and produce the 3-D DWT results. Fig.~\ref{blockdia_1}(b) shows that there is no requirement of temporal buffer, due to the sub-band coefficients of two spatial processors are directly connected to the four temporal processors. But it has been designed with 2 pipeline stages, it require 8 pipeline registers for each TP.  Same frequency sub-band of the distinct spatial processors are fed to the each temporal processor. i.e. LL, HL, LH and HH sub-bands of spatial processor 1 and 2 are given as inputs to the temporal processor 1, 2, 3 and 4 respectively. Temporal processor apply 1-D Haar wavelet on sub-band coefficients, and provide the low frequency sub-band and high frequency sub-band as output. By combining all low frequency sub-bands and high frequency sub-bands of all temporal processors provide the 3-D DWT output in the form of L-Frame and H-Frame (2-D DWT by spatial processors and 1-D DWT by temporal processors). 

\section{Architecture for Compressed Sensing Module}
\begin{figure} 
\centering
\includegraphics [height=80mm,width=90mm]{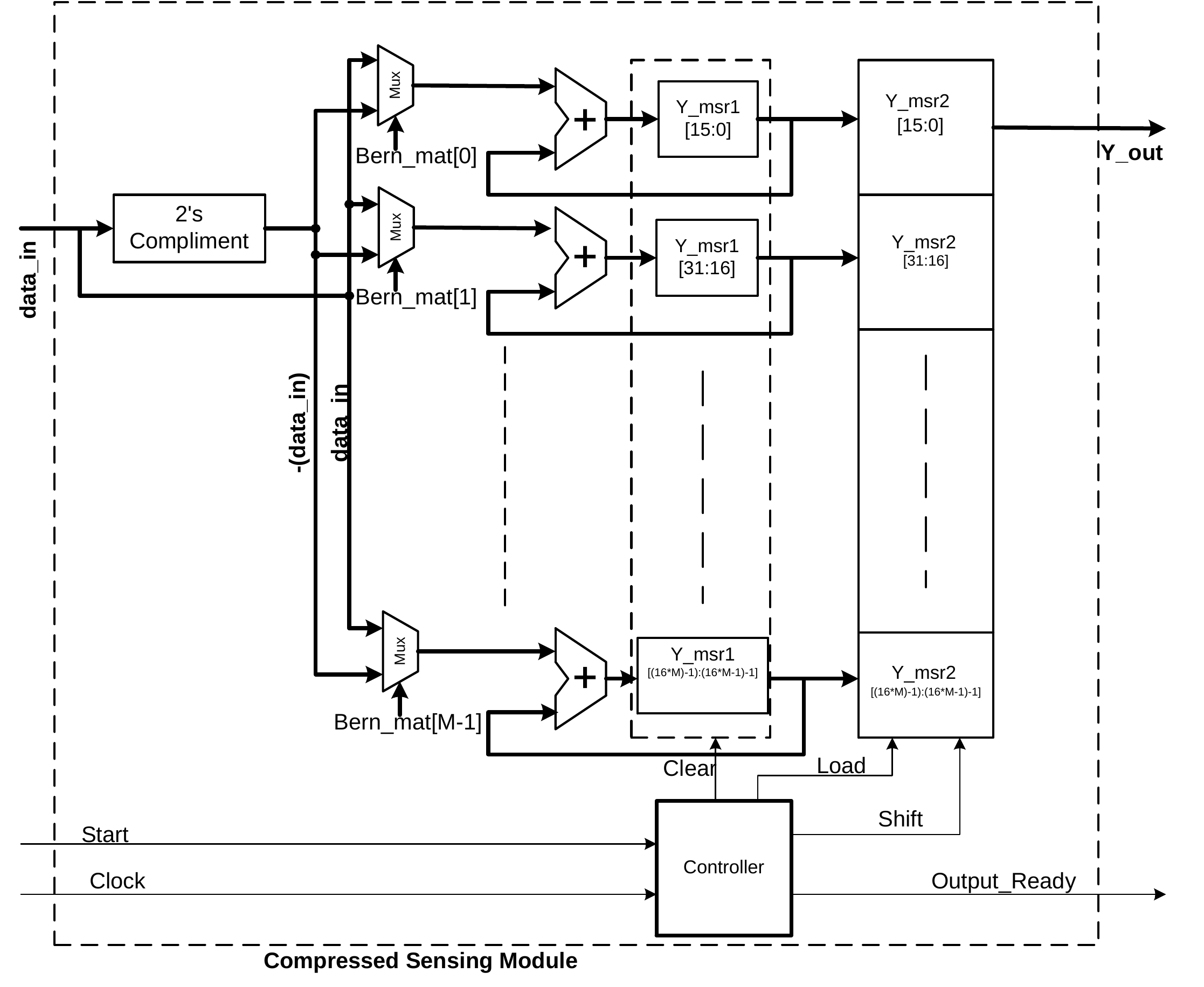}
\caption{Internal architecture of CS module}
\label{cs_mod}
\end{figure}
\par The proposed 3-D DWT module, simultaneously works on two video frames of size $ N\times N $ and  provide eight 3-D DWT sub-bands as its output. As shown in Fig. \ref{blockdia_1}(b), CS is applied on all sub-bands of 3-D DWT outputs, except LLL band (LL band of L-Frame) and each sub-band is connected to one CS module. Size of the each sub-band equals to the half of the original frame for one level decomposition (N/2$ \times $N/2). The main function of the CS module is to calculate the measured matrix $ y $ from $ \Phi $ and $x$ by using the CS equation $y = \Phi x$. Where $ x $ is a input vector (for which CS need to calculate). Size of $ x $ is equal to P* N/2 (N/2 is the height of single column in a sub-band), because proposed 3-D DWT simultaneously works on P columns due to P number of PUs in the spatial processor.  Proposed architecture has been designed with P = 2; so for each clock,  alternative column coefficients are provided by the 3-D DWT module for each sub-band. With P equals to 2, the size of $ x $ is [2*N/2]$ \times $1  = N$ \times $1, and $ \Phi $ is the randomly generated Bernoulli matrix, size of the $\Phi$ is M$ \times $N, ($M \ge cK\log (N/K)$ for some small constant c). Value of  $ K $ ($ \ell_0 $-norm) of the input vector $ x $.  We have tested for different video sequences of size 512$ \times $512 and 1024$ \times $1024 with different threshold values (wavelet coefficient value less than the threshold value consider as zero) have been observed and it shows that the value of K is not more than N/8 for given $ x $ of size  N$ \times $1. Based on those observations, value of $ M $ has been fixed to N/4. 
\par Fig. \ref{cs_mod} shows the internal architecture of CS module. Proposed architecture for CS based encoder has seven CS modules, one for each sub-band except LLL sub-band. The structure of seven CS modules are same and works simultaneously. For all these seven CS modules only one Bernoulli matrix has been used and it is stored in ROM, denoted by Bern$\_$mat. The size of the Bern$\_$mat is M$ \times $N, each location has M bits representing one entire column and number of locations equals to N. Bernoulli matrix has been generated by using `binord' function in the Matlab tool ($\Phi$ = binornd(1,0.5,M,N)), with equal probability for 0 and 1 of size M$ \times $N. Here bit `0' represents the value `+1' and 1 represents `-1'. This generated Bernoulli matrix has been loaded in the Bern$\_$mat (ROM) locations and is used by all CS modules. As shown in Fig. \ref{cs_mod}, input for a CS module is data$ \_ $in which is sub-band out from 3-D DWT. For every clock one 15-bit data$ \_ $in will arrive (alternative column per each clock). In $y = \Phi x$, $ y $ is column matrix of size M$ \times $1, which is represented as $ y $ = [$y_{0}$, $y_{1}$, $y_{2}$,  $y_{3}$, ........ $y_{M-1}$]$^{T}$, $y_{i}$=$\sum\limits_{k = 1}^N {{\Phi _{ik}}{x_k}}$ or we can also calculate iterative fashion for every $(n+1)^{th}$ clock $y_{i}(n+1)$ = $y_{i}(n)$ + $\Phi _{ik}{x_k}(n+1)$, it require N clocks to complete this operation, because $ k $ = 0 to N-1.
\par The proposed architecture uses M adders, one for each individual measurement $y_{i}$. One input of the adder is $\Phi _{ik}{x_k}$ which is the output of a multiplexer, where $\Phi _{ik}$ is either 0 or 1, if $\Phi _{ik}$ is 0, then $ x_{k}$ multiply with +1 ($\Phi _{ik}x_{k} = $data$ \_ $in), otherwise multiply with -1 ($\Phi _{ik}x_{k}$ = 2's compliment of (data$ \_ $in )), this task has been done by connecting the $\Phi _{ik}$ as a selection line of the multiplexer and first and second inputs of the multiplexer is $x_{k}$ and $- x_{k}$ respectively. The second input of adder is from partial result of $y_{i}$ in the previous clock. The proposed architecture for CS module utilize the two registers to store the M measurements ($y$) namely, Y$ \_ $msr1 and Y$ \_ $msr2, each of capacity M*16 bits (16 bits for each measurement). Y$ \_ $msr1 is used store partial results of $y_{i}$ from 0 to N-1 clocks. Just after completing N clocks, measurements are ready and are available in Y$ \_ $msr1, then control circuit transfer the Y$ \_ $msr1 data to Y$ \_ $msr2 and clear the Y$ \_ $msr1 for next set of measurements. The above procedure is repeated for all the columns of sub-band at the same time calculated measurements $y_{i}$ each of 16-bit are send as output (Y$\_ $out) from Y$\_ $msr2 by  shifting 16 bits for each clock. This procedure is followed for all the seven sub-bands. Each measured matrix $ y $ is sent for the entropy coding (Golomb Rice Coding) block and coded bit streams are transmitted through channel. LLL sub-band is directly coded by entropy coding block and then transmitted through channel by considering as a base layer. Entropy coding is out of scope of this paper, not discussed in this paper. 

\section{Results and Performance Comparison} 
\subsection{Simulation Results}
The proposed encoder has been simulated by using Matlab tool and functionality has been verified on $ cyclone $ (Downloaded from the NASA website) and $ clock $ video sequences of 512$ \times $512 resolution, $ viplane $ and $ foreman $ video sequences of 256$ \times $256 resolution. After applying the 3-D DWT, all the HL, LH and HH sub-bands of L-Frames and LL, HL, LH and HH sub-bands of H-Frames are sent to CS.  After applying the CS on 3-D DWT coefficients measurements are passed through the entropy coder (Golomb Rice coding + run length encoding). Percentage of measurements are calculated before entropy coding. Compression Ration is the ratio of total number of bits in input frame and number of bits after the entropy coding. Table \ref{table_1} shows that performance of the proposed framework competes with the existing IBMCTF \cite{CS_4} and H.264 \cite{H.264}. Performance in terms of compression ratio and PSNR of the proposed encoder and decoder for $ clock, cyclone $ and $ Viplane $ video sequences are noted from the level 1 to level 3 in Table \ref{table_2}.
\begin{table}
\begin{center}
\caption{Performance  of Proposed Framework with IBMCTF and H.264}
\label{table_1}
\begin{tabular}{|l|l|c|c|}
\hline
              CODEC                                               & Video     & Compression Ratio & PSNR (dB)  \\ \hline
\multirow{3}{*}{Proposed}                  & clock     & 24.24            & 44.01 \\ \cline{2-4} 
                                                             & cyclone & 16.85            & 34.2  \\ \cline{2-4} 
                                                             & viplane   & 20.96            & 37.5  \\ \hline
\multirow{3}{*}{IB-MCTF \cite{CS_4}} & clock     & 7.33             & 46.2  \\ \cline{2-4} 
                                                             & cyclone & 5.08             & 40.6  \\ \cline{2-4} 
                                                             & viplane   & 5.28             & 47.33 \\ \hline
\multirow{3}{*}{H.264 \cite{H.264}}                                       & clock     & 62.33            & 42.65 \\ \cline{2-4} 
                                                             & cyclone & 22.1             & 38.4  \\ \cline{2-4} 
                                                             & viplane   & 37.8             & 40.57 \\ \hline
\end{tabular}
\end{center}
\end{table}

\begin{table*}
\begin{center}
\caption{Performance of the proposed framework for different video sequences and different levels}
\label{table_2}
\begin{tabular}{|l|c|c|c|c|}
\hline
\multirow{2}{*}{Video clip}                                                                  & \multirow{2}{*}{level} & \multirow{2}{*}{PSNR} & \multirow{2}{*}{Compression Ratio} & \multirow{2}{*}{\begin{tabular}[c]{@{}l@{}}\% of measurements \\ before Entropy coding\end{tabular}} \\
                                                                                             &                        &                       &                                    &                                                                                         \\ \hline
\multirow{3}{*}{\begin{tabular}[c]{@{}l@{}}Clock\\  512x512\\ (Slow motion)\end{tabular}}    & 1                      & 44                    & 24.24                              & 34.99                                                                                   \\ \cline{2-5} 
                                                                                             & 2                      & 33.2                  & 41.67                              & 23.82                                                                                   \\ \cline{2-5} 
                                                                                             & 3                      & 30.12                 & 53.23                              & 20.52                                                                                   \\ \hline
\multirow{3}{*}{\begin{tabular}[c]{@{}l@{}}Cyclone\\ 512x512\\ (High motion)\end{tabular}} & 1                      & 34                    & 16.85                              & 43.7                                                                                    \\ \cline{2-5} 
                                                                                             & 2                      & 29                    & 20.56                              & 38.61                                                                                   \\ \cline{2-5} 
                                                                                             & 3                      & 25.5                  & 23.3                               & 36.6                                                                                    \\ \hline
\multirow{3}{*}{\begin{tabular}[c]{@{}l@{}}Viplane\\ 256x256\\ (Medium motion)\end{tabular}} & 1                      & 37.5                  & 20.96                              & 32.7                                                                                    \\ \cline{2-5} 
                                                                                             & 2                      & 31.5                  & 35.63                              & 23.5                                                                                    \\ \cline{2-5} 
                                                                                             & 3                      & 28                    & 65.54                              & 18.12                                                                                   \\ \hline
\end{tabular}
\end{center}
\end{table*}

\subsection{Synthesis Results}
The proposed architecture for CS based low complex video encoder has been described in Verilog HDL. Simulation results have been verified by using Xilinx ISE simulator. We have simulated the Matlab model which is similar to the proposed CS based low complex video encoder architecture and verified the 3-D DWT coefficients and CS measurements. RTL simulation results have been found to exactly match the Matlab simulation results. The Verilog RTL code is synthesised using Xilinx ISE 14.2 tool and mapped to a Xilinx programmable device (FPGA) 7z020clg484 (zync board) with speed grade of -3. Table \ref{FPGA_results} shows the device utilisation summary of the proposed architecture and it operates with a maximum frequency of 265 MHz. The proposed architecture has also been synthesized using SYNOPSYS design compiler with 90-nm technology CMOS standard cell library. Synthesis results of the proposed encoder is provided in Table \ref{enc_synopsys}, it consumes 90.08 mW power and occupies an area equivalent to 416.799 K equivalent gate count at frequency of 158 MHz.
 
\begin{table}
\centering
\caption{Device utilization summary of the proposed Encoder}
\label{FPGA_results}
\begin{tabular}{|l|c|c|c|}
\hline
Logic utilized       & Used                  & Available             & Utilization (\%)      \\ \hline
Slice Registers      & 15917                  & 106400                & 14\%                   \\ \hline
Number of Slice LUTs & 47303                  & 53200                 & 88\%                   \\ \hline
Number of fully      & \multirow{2}{*}{15523} & \multirow{2}{*}{47697} & \multirow{2}{*}{32\%} \\ 
used LUT-FF pairs    &                       &                       &                       \\ \hline
Number of Block RAM  & 3                     & 140                   & 2\%                   \\ \hline
\end{tabular}
\end{table}
\begin{table}
\centering
\caption{Synthesis Results for Proposed Encoder}
\label{enc_synopsys}
\begin{tabular}{|l|c|}
\hline
Combinational Area     & 1072673 $ \mu m^{2} $ \\ \hline
Non Combinational Area & 915778 $ \mu m^{2} $  \\ \hline
Total Cell Area        & 1988451 $ \mu m^{2} $ \\ \hline
Interconnect area      &  316449 $ \mu m^{2} $ \\ \hline
Operating Voltage      & 1.2 V                 \\ \hline
Total Dynamic Power    & 80.17 mW              \\ \hline
Cell Leakage Power     & 9.90 mW               \\ \hline
\end{tabular}
\end{table}

\subsection{Comparison}

Table \ref{3dcompare} shows the comparison of proposed 3-D DWT architecture with existing 3-D DWT architecture. It is found that, the proposed design has less memory requirement, High throughput, less computation time and minimal latency compared to \cite{3D_weeks}, \cite{3D_tagavi}, \cite{3D_swapna}, and \cite{3D_darji}. Though the proposed 3-D DWT architecture has small disadvantage in area and frequency, when compared to \cite{3D_swapna}, the proposed one has a great advantage in remaining all aspects.   
 
 Table \ref{3d_asic} gives the comparison of synthesis results between the proposed 3-D DWT architecture and \cite{3D_darji}. It seems to be proposed one occupying more cell area, but it included total on chip memory also, where as in \cite{3D_darji} on chip memory is not included. Power consumption of the proposed 3-D architecture is very less compared to \cite{3D_darji}.

\begin{table}
\centering
\caption{Comparison of proposed 3-D DWT architecture with existing architectures (for 1-level)}
\label{3dcompare}
\resizebox{\textwidth}{!}
{\begin{tabular}{|l|l|l|l|l|l|}
\hline
Parameters   & Weeks \cite{3D_weeks}        & Taghavi \cite{3D_tagavi} & A.Das  \cite{3D_swapna} & Darji \cite{3D_darji} & Proposed        \\ \hline
\hline
Memory requirement & $6N^{2} $+$ 6l$         & $5N^{2} $                   & $5N^{2} $   + 5N           & $4N^{2} $   + 10N        & 2*(3N+40P)              \\ \hline
Throughput/cycle   & -                               & 1 result                    & 2 results                  & 4 results                & 8 results       \\ \hline
Computing time & \multirow{2}{*}{$2N^{2}$ +  $ 3l$/2 }        & \multirow{2}{*}{$6N^{2} $} & \multirow{2}{*}{$3N^{2} $}             & \multirow{2}{*}{$3N^{2} $}           & \multirow{2}{*}{ $N^{2}$/2P } \\  For 2 Frames &            &                                  &           &        &             \\ \hline
Latency            & $2.5N^{2} $ + 0.5$ l $   & $4N^{2} $ cycles            & $2N^{2} $ cycles           & $3N^{2} $/2 cycles       & 12 cycles       \\ \hline
Area               & -                               & -                           & 1825 slices                & 2490 slices              & 2852 slice LUTs \\ \hline
Operating          & \multirow{2}{*}{200 MHz (ASIC)} & \multirow{2}{*}{-}          & 321 MHz                    & 91.87 MHz                & 265 MHz         \\ 
Frequency          &                                 &                             & (FPGA)                     & (FPGA)                   & (FPGA)          \\ \hline
Multipliers        & -                               & -                           & Nil                        & 30                       & Nil             \\ \hline
Adders             & $ 6l$ MACs                      & -                           & 78                         & 48                       &      176           \\ \hline
Filter bank        & $l$-length                      & D-9/7                       & D-9/7                      & D-9/7                    & D-9/7 (2-D) + Haar (1-D)           \\ \hline
\end{tabular}}
\end{table}

\begin{table}
\centering
\caption{Synthesis Results (Design Vision) Comparison of Proposed 3-D DWT architecture with existing}
\label{3d_asic}
\begin{tabular}{|l|c|c|}
\hline
Parameters             & Darji et al.,\cite{3D_darji}     & Proposed    \\ \hline
Combinational Area     & 61351 $ \mu m^{2} $  & 526419 $ \mu m^{2} $  \\ \hline
Non Combinational Area & 807223 $ \mu m^{2} $ & 553078 $ \mu m^{2} $  \\ \hline
Total Cell Area        & 868574 $ \mu m^{2} $ & 1079498 $ \mu m^{2} $ \\ \hline
Operating Voltage      & 1.98 V     & 1.2 V        \\ \hline
Total Dynamic Power    & 179.75 mW  & 38.56 mW    \\ \hline
Cell Leakage Power     & 46.87 $ \mu W $   & 4.86 mW     \\ \hline
\end{tabular}
\end{table}
 
\section{Conclusions}
In this paper, we have proposed memory efficient and high throughput architecture for CS based low complex encoder. The proposed architecture is implemented on 7z020clg484 FPGA target of zync family, also synthesized on Synopsys' design vision for ASIC implementation. An efficient design of 2-D spatial processor and 1-D temporal processor reduces the internal memory, latency, CPD and complexity of a control unit, and increases the throughput. When compared with the existing architectures the proposed scheme shows higher performance at the cost of slight increase in area. The proposed encoder architecture is capable of computing 60 UHD (3840$ \times $2160) frames in a second. The proposed architecture is also suitable for scalable video coding. In addition, the  complexity of the encoder is reduced to a great extent. The proposed encoder is considered to be suitable for applications including satellite communication, wireless transmission and data compression by high speed cameras.

\end{document}